\documentclass[iop]{emulateapj}
\slugcomment{{\sc Accepted for publication:} September 5, 2011}

\usepackage{rotating}
\usepackage{threeparttable}
\usepackage{color} 
\newcommand{\red}[1]{\textcolor{black}{#1}}

\shorttitle{The ultraluminous X-ray source NGC~5408~X-1}
\shortauthors{Gris\'e et al.}

\begin{document}

\title{Optical emission of the ultraluminous X-ray source NGC~5408~X-1: donor star or irradiated accretion disk?}

\author{F. Gris\'e\altaffilmark{1}, P. Kaaret\altaffilmark{1}, S. Corbel\altaffilmark{2}, H. Feng\altaffilmark{3}, D. Cseh\altaffilmark{2}, L. Tao\altaffilmark{3}}

\altaffiltext{1}{Department of Physics and Astronomy, University of
Iowa,  Van Allen Hall, Iowa City, IA 52242, USA; \email{fabien-grise@uiowa.edu}}
\altaffiltext{2}{Laboratoire Astrophysique des Interactions Multi-\'echelles (UMR 7158), CEA/DSM-CNRS-Universit\'e Paris Diderot, CEA Saclay, F-91191 Gif sur Yvette, France}
\altaffiltext{3}{Department of Engineering Physics and Center for Astrophysics, Tsinghua University, Beijing 100084, China}

\begin{abstract}
We obtained 3 epochs of simultaneous {\it Hubble Space Telescope (HST)}/Wide Field Camera 3 and {\it Chandra} observations of the ultraluminous X-ray source NGC~5408~X-1. The counterpart of the X-ray source is seen in all {\it HST} filters, from the UV through the NIR, and for the first time, we resolve the optical nebula around the ULX. We identified a small OB association near the ULX that may be the birthplace of the system. The stellar association is young, $\sim 5\ \mathrm{Myr}$, contains massive stars up to $40\ \mathrm{M_{\odot}}$, and is thus similar to associations seen near other ULXs, albeit younger. The UV/Optical/NIR spectral energy distribution (SED) of the ULX counterpart is consistent with that of a B0I supergiant star. We are also able to fit the whole SED from the X-rays to the NIR with an irradiated disk model. The three epochs of data show only marginal variability and thus, we cannot firmly conclude on the nature of the optical emission.

\end{abstract}


\section{Introduction}

Ultraluminous X-ray sources (ULXs) are extragalactic sources that are not at the nucleus of their galaxy and emit well above the Eddington limit of a $20\ \mathrm{M_\odot}$ black hole ($L_X \sim 3\times 10^{39}\ \mathrm{erg \; s^{-1}}$). The most important, still unresolved question is whether these objects contain intermediate mass black holes - IMBHs - ($M > 100\ \mathrm{M_\odot}$) or stellar mass black holes with super-Eddington or beamed emission (see \citealt{Feng11} for a review).

A direct answer regarding the mass of compact objects in ULXs would be possible if a radial velocity curve of absorption lines from the companion star, or emission lines coming from the accretion disk, could be obtained. To date, this has been proven to be a real challenge due to the lack of absorption lines in ULX optical spectra and what seem to be non-periodic variations of the only traceable emission line\red{, \ion{He}{2}$\lambda$4686,} in Holmberg~IX~X-1 and NGC~1313~X-2 \citep{Roberts10}. The only exception, to date, is the optically very luminous source P13 in NGC~7793 \citep{Motch10} where absorption lines from a late B supergiant star (in addition to a \ion{He}{2} emission line) have been observed. This is a very promising candidate to constrain the mass function of the system. However, this is the only example out of more than ten identified optical counterparts.

In all other ULXs, the true nature of the donor star is mostly unknown. The X-ray lightcurve of M82~X-1 shows a 62 day periodicity consistent with a strictly periodic signal and hence is likely the orbital period of the binary system \citep{Kaaret07}. If the companion star fills its Roche lobe, this means that such star has to be of low-density and hence would be a giant or supergiant star \citep{Kaaret06}. Other constraints on the nature of the donor stars rely on studies of their environment. Indeed, some counterparts belong to or are located nearby small stellar associations (\citealt{Soria05,Grise08,Grise11}), where it is possible to constrain the association age, and thus the maximum mass for a member star. This has led to maximum mass estimates of $15-20\ \mathrm{M_{\odot}}$ that apply as well to the donor star of the ULX systems, if mass transfer did not play an important role. In some other cases \citep{Feng08,Yang11}, the constraints arise from field stars in the absence of such associations. But these estimates are based on indirect evidence and the precise stellar type of those stars remains unknown.

The most obvious problem in identifying the companion star in the optical emission of ULXs is that the flux contribution coming from the accretion disk (direct and irradiated) is not known but is definitely present in some sources. The best hint comes from NGC~1313~X-2, where its optical lightcurve is contaminated by short-term, stochastic variations that are likely due to X-ray reprocessing in the accretion disk \citep{Grise08,Liu09,Impiombato10} and also from X-ray heating in the companion star \citep{Zampieri11}. Broad \ion{He}{2} emission lines with equivalent widths up to ten times higher than in the brightest Galactic high-mass X-ray binaries are present in the optical spectrum of some counterparts (e.g. \citealt{Pakull06,Roberts10,Cseh11}) and are also a direct proof of significant X-ray reprocessing in these systems.

The counterpart optical colors have often been used to place constraints on the companion star.  The high optical/UV luminosities and blue colors suggest early-type, OB stars (e.g. \citealt{Liu02, Kaaret04, Soria05}). But, as suggested by \citet{Kaaret05} and \citet{Pakull06}, the optical light may be contaminated, or even dominated, by the irradiated accretion disk (\citealt{Kaaret09,Tao11}). This is similar to active low-mass X-ray binaries, although ULXs have higher X-ray luminosities and may have more massive donor stars. Unfortunately, the optical spectral energy distributions of hot, irradiated disks look rather like those of OB stars.

Our present work is devoted to a well studied ULX, NGC~5408~X-1 (hereafter N5408X1), \red{located at 4.8 Mpc \citep{Karachentsev02}}. N5408X1 is a bona-fide ULX with an average luminosity (0.3-10 keV) of $1\times 10^{40}\ \mathrm{erg \; s^{-1}}$ \citep{Strohmayer09} that displays clear variability within a factor of $\sim 2$ \citep{Kaaret09b,Strohmayer09,Kong11}. Intensive {\it Swift} monitoring of the source has been used to infer a $\sim 115\ \mathrm{day}$ period that has been interpreted as the orbital period of the ULX system \citep{Strohmayer09}, although \citet{Foster10} suggests that this modulation may instead be superorbital. The X-ray spectrum of N5408X1 does not seem to vary much \citep{Kaaret09b}. All observations performed to date show a soft spectrum $\Gamma \sim 2.5$ and the presence of a soft component. Different interpretations have been put forward: if interpreted in the framework of Galactic black holes binaries (GBHBs), standard models (powerlaw plus disk component) point to the presence of a massive black hole if the temperature of the disk scales as in Galactic counterparts \citep{Kaaret03,Soria04,Strohmayer07}. The power-law would be in that case the signature of a corona as seen in GBHBs. Instead, the interpretation in the context of super-critical accreting systems would mean that the black hole is much smaller, with the possibility that a strong wind is the origin of the soft component, while the high energy part of the spectrum would be due to disk emission inside the spherization radius \citep{Poutanen07}. Alternatively, the soft component may represent emission from the outer disk with the hard emission associated with thermal Comptonization \citep{Gladstone09,Middleton11}. \red{Finally, the soft excess may arise from reflection from the disk \citep{Caballero10}.}

N5408X1 is also one of the few ULXs where quasi-periodic oscillations (QPOs) have been discovered \citep{Strohmayer07,Strohmayer09}. By analogy with Galactic black hole binaries (GBHBs), \citet{Strohmayer09} have interpreted this feature as a type C low-frequency QPO (LFQPO), which would indicate that N5408X1 hosts an IMBH. \citet{Middleton11} have challenged this conclusion and concluded that even the most super-Eddington BHB, GRS 1915+105, is not really a good match to the properties of N5408X1. They hypothesized that the temporal and spectral properties are more consistent with the feature seen in \red{some} Narrow Line Seyfert 1 galaxies which are supposed to be even higher super-Eddington sources, which would indicate a black hole with mass $\la 100\ \mathrm{M_{\odot}}$. But the large uncertainties associated with the black hole masses in those systems avoids any definitive conclusion.

N5408X-1 is also seen at other wavelengths. This source is one of the few ULXs detected in radio, the emission being consistent with a powerful nebula surrounding the ULX \citep{Kaaret03,Soria06,Lang07,Cseh11b}. This nebula is also detected in optical wavelengths \citep{Pakull03,Soria06,Kaaret09} where the optical emission is consistent with X-ray photoionization. The optical counterpart of the ULX, first uniquely identified in a {\it Hubble Space Telescope (HST)}/WFPC2 image by \citet{Lang07} as a $V_{0}=22.2$ magnitude object, has been subsequently studied using VLT spectroscopy by \citet{Kaaret09} and \citet{Cseh11}. The optical spectrum of N5408X1 is similar to many other ULXs, displaying a blue continuum with superimposed strong and narrow emission lines coming from the nebula. Of prime interest is the presence of broad components with widths of $\sim 750\ \mathrm{km \; s^{-1}}$ in the \ion{He}{2} and $H_\beta$ lines \citep{Cseh11} that appear spatially point-like, similar to the optical continuum of the counterpart. The radial velocity of the broad \ion{He}{2} line varies, which may represent the motion of the compact object or motions within the accretion disk (see \citealt{Roberts10} for the behavior of this line in other sources).

In this paper, we present new, simultaneous {\it Chandra}/{\it HST} observations of N5408X1. {\it HST} observations that allow us to present for the first time the spectral energy distribution of a ULX from the ultraviolet (UV) to the near-infrared (NIR) wavelengths. Combined with simultaneous X-ray data, we are able to show that an irradiated disk model fits well all the data. We discuss other alternatives and suggest a way to confirm this result with new observations.

\section{Observations and data analysis}

We obtained three series of multiwavelength observations using the recently installed Wide Field Camera 3 (WFC3) instrument aboard the {\it Hubble Space Telescope (HST)} (Program ID 12010, PI: P. Kaaret, along with simultaneous {\it Chandra} X-ray observations (Program ID 11400085, PI: P. Kaaret).

\subsection{Optical observations}

Each {\it HST} visit consisted of observations in UV, optical, and near-infrared filters. UV and optical observations were carried out in the F225W, F336W, F547M, and F845M filters using the WFC3/UVIS camera. We used the WFC3/IR camera for the near-infrared observations, in the F105W and F160W filters. A single observation was done with the WFC3/UVIS using the narrow F502N filter to isolate emission due to the forbidden [\ion{O}{3}]$\lambda$5007 line. A summary of the observations can be found in Table \ref{obs}.

We performed aperture and point-spread function (PSF) photometry on the drizzled images.
Aperture photometry was done only on the four UVIS images, since the crowding is really severe in the infrared images, preventing any reliable photometry in the region nearby the ULX with this method. We performed the aperture photometry with Sextractor v. 2.5.0 \citep{Bertin96}, using an aperture of 2 pixels in radius (i.e, 0.08\arcsec). The background was estimated by relying on the global mapping done by Sextractor. Aperture corrections were calculated using relatively bright and isolated stars, performing photometry using a 0.4\arcsec\ radius (i.e. 10 pixels).

PSF photometry was performed using {\small DAOPHOT~II} \citep{Stetson87} under {\small ESO-MIDAS}.
For the UVIS images, we used a 2.5 pixel aperture (i.e. $0.10\arcsec$) which corresponds approximately to 60-70\% of the enclosed energy \citep{Kalirai09b} depending on the filter. The background was chosen to be a 10 pixel width annulus located at 15 pixels from the object centroid. Given the use of a sub-array for the UVIS observations, we used a constant PSF across the frame.  Aperture corrections were calculated using the aperture photometry routine within the {\small DAOPHOT package} in the same way as previously described.
A similar procedure was applied to the IR images, with a 2.0 pixel aperture (i.e. $0.26\arcsec$), corresponding to 70-80\% of the enclosed energy \citep{Kalirai09a}. The IR observations were performed in full frame mode, thus we used a quadratically varying PSF in the field. Aperture corrections were measured as for the UVIS images.
Finally, zeropoints calculated for a $0.4\arcsec$ radius were taken out of \citet{Kalirai09b,Kalirai09a} for UVIS and IR observations.

It turns out that the main uncertainty in the photometry is due to the aperture correction. Indeed, the use of a window for the UVIS observations means that the number of bright stars is low which forces us to rely on fewer stars, or less bright stars, for the correction.

In the rest of this paper, we will use the results coming from the PSF photometry, but the two methods give consistent results for bright or isolated stars, with differences below 0.1 mag.

\begin{table*}[!t]
\caption[]{The {\it HST}/WFC3 observations for NGC~5408~X-1.}
\label{obs}
\centering
\begin{tabular}{cccccc}
\hline
\hline
	&ID & Instrument & Filter & Date  & Exposure time (s)\\
\hline
Epoch 1 &ibde01030 & UVIS &  F225W  & 2010 May 02 	 &     560  \\
	&ibde01020 & UVIS &  F336W  & 2010 May 02 	 &     280  \\
	&ibde01010 & UVIS &  F547M  & 2010 May 02 	 &     200  \\
	&ibde01040 & UVIS &  F845M  & 2010 May 02 	 &     280  \\
	&ibde01060 & IR   &  F105W  & 2010 May 02 	 &     298.5  \\
	&ibde01050 & IR   &  F160W  & 2010 May 02 	 &     498.5  \\
\hline
\hline
Epoch 2	&ibde02030 & UVIS &  F225W  & 2010 May 15	&     560  \\
	&ibde02020 & UVIS &  F336W  & 2010 May 15	&     280  \\
	&ibde02010 & UVIS &  F547M  & 2010 May 15 	&     200  \\
	&ibde02040 & UVIS &  F845M  & 2010 May 15 	&     280  \\
	&ibde02060 & IR   &  F105W  & 2010 May 15 	&     298.5  \\
	&ibde02050 & IR   &  F160W  & 2010 May 15 	&     498.5  \\
\hline
\hline
Epoch 3	&ibde53030 & UVIS &  F225W  & 2010 Sep. 12 	&     560  \\
	&ibde53020 & UVIS &  F336W  & 2010 Sep. 12 	&     280  \\
	&ibde53010 & UVIS &  F547M  & 2010 Sep. 12 	&     200  \\
	&ibde53040 & UVIS &  F845M  & 2010 Sep. 12 	&     280  \\
	&ibde53060 & IR   &  F105W  & 2010 Sep. 12 	&     298.5  \\
	&ibde53050 & IR   &  F160W  & 2010 Sep. 12 	&     498.5  \\
\hline
\hline
	&ibde04010 & UVIS &  F502N  & 2010 Dec. 26 	&     2600  \\
\hline

\end{tabular}
\end{table*}

\begin{table*}
\caption[]{Brightness of the ULX counterpart in different filters, from the {\it HST}/WFC3 observations. Magnitudes are expressed in the {\it HST}/WFC3 Vegamag system. The errors on the magnitude include the aperture correction uncertainty. The value between brackets is the photometric error, as derived by {\small DAOPHOT}.}
\label{tab_magnitudes_hst_hoix}
\centering
\small
\begin{tabular}{cccccccc}
\hline
\hline
	& Filter &  Aperture correction & VEGAmag & Flux ($10^{-18}\ \mathrm{erg \; s^{-1} \; cm^{-2} \; \AA^{-1}}$)\\
\hline
Epoch 1 &F225W      & 0.44 $\pm$ 0.03	& 20.09 $\pm$ 0.04 (0.024)  & $38.60 \pm 1.42$	\\
	&F336W      & 0.42 $\pm$ 0.06	& 20.64 $\pm$ 0.07 (0.029)  & $18.02 \pm 1.16$	\\
	&F547M      & 0.53 $\pm$ 0.06	& 22.43 $\pm$ 0.07 (0.029)  & $3.91 \pm 0.25$	\\
	&F845M      & 0.59 $\pm$ 0.04	& 22.42 $\pm$ 0.08 (0.067)  & $1.03 \pm 0.08$	\\
	&F105W      & 0.06 $\pm$ 0.03	& 22.84 $\pm$ 0.09 (0.072)  & $0.34 \pm 0.03$	\\
	&F160W      & 0.20 $\pm$ 0.05	& 22.48 $\pm$ 0.10 (0.091)  & $0.12 \pm 0.01$	\\
\hline
\hline
Epoch 2	&F225W      & 0.42 $\pm$ 0.04	& 20.04 $\pm$ 0.05 (0.034)  & $40.42 \pm 1.86$	\\
	&F336W      & 0.45 $\pm$ 0.05	& 20.55 $\pm$ 0.05 (0.022)  & $19.58 \pm 0.90$	\\
	&F547M      & 0.50 $\pm$ 0.06	& 22.30 $\pm$ 0.07 (0.026)  & $4.41 \pm 0.28$	\\
	&F845M      & 0.54 $\pm$ 0.04	& 22.37 $\pm$ 0.07 (0.052)  & $1.08 \pm 0.07$	\\
	&F105W      & 0.22 $\pm$ 0.07	& 22.63 $\pm$ 0.10 (0.090)  & $0.41 \pm 0.04$	\\
	&F160W      & 0.22 $\pm$ 0.05	& 22.37 $\pm$ 0.10 (0.073)  & $0.14 \pm 0.01$	\\
\hline
\hline
Epoch 3	&F225W      & 0.49 $\pm$ 0.04	& 19.97 $\pm$ 0.04 (0.028)  & $43.11 \pm 1.59$ 	\\
	&F336W      & 0.53 $\pm$ 0.08	& 20.56 $\pm$ 0.08 (0.021)  & $19.40 \pm 1.43$	\\
	&F547M      & 0.51 $\pm$ 0.08	& 22.26 $\pm$ 0.08 (0.014)  & $4.58 \pm 0.34$\\
	&F845M      & 0.61 $\pm$ 0.04	& 22.26 $\pm$ 0.06 (0.038)  & $1.20 \pm 0.07$	\\
	&F105W      & 0.16 $\pm$ 0.07	& 22.55 $\pm$ 0.12 (0.076)  & $0.44 \pm 0.05$	\\
	&F160W      & 0.14 $\pm$ 0.06	& 22.44 $\pm$ 0.10 (0.100)  & $0.13 \pm 0.01$	\\
\hline
\end{tabular}
\normalsize
\end{table*}

\subsection{X-ray observations}

N5408X1 was observed with {\it Chandra} using the {\it Advanced CCD Imaging Spectrometer} (ACIS) instrument. For each series of {\it HST} observations, we obtained a simultaneous 12 ks {\it Chandra} exposure. Given the high count rate expected of N5408X1, only one chip (ACIS S3) was used, with a 1/8 subarray mode. This reduces the frame time to 0.441 s which mitigates pile-up in the data. \red{The observed count rate in the three observations varies from $\sim 0.3$ to $\sim 0.4$~counts/s.}
Standard extraction of spectra was done using the CIAO 4.3 package \citep{Fruscione06} with CALDB 4.4.1. We rebinned the spectra to a minimum of a $5 \sigma$ significance per bin. All fitting was performed using {\small ISIS} v.1.6.1-26 \citep{Houck00}.

\section{Results}

\begin{figure}[tb]
\resizebox{8cm}{!}{\includegraphics{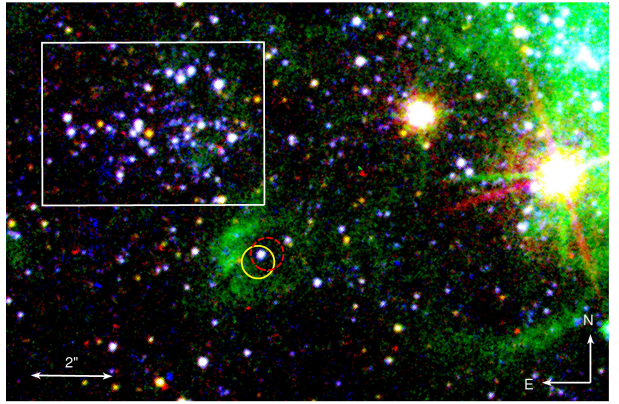}}
\caption[]{{\it HST}/WFC3 composite image (blue - F225W, green - F502N, red - F845M) of the field surrounding NGC~5408~X-1. $1\arcsec$ represents 23.3 pc at the distance of NGC~5408~X-1.
Overplotted are the radio positions from VLA (yellow full circle) and ATCA observations (red dashed circle) \citep{Lang07} with a conservative error circle of $0.4\arcsec$ in radius.
The white box ($5.5\arcsec \times 4.0 \arcsec$) shows the nearby stellar association. Stars located inside this rectangle were plotted differently in the color-magnitude diagrams (see Figures \ref{cmd} \& \ref{et}).}
\label{image}
\end{figure}

The optical counterpart of the ULX, as identified by \citet{Lang07}, is visible in all the {\it HST} filters used in this study. Figure \ref{image} shows a composite image including the F225W, F502N ([\ion{O}{3}]) and F845M filters. Magnitudes and corresponding fluxes of the ULX counterpart are presented in Table \ref{tab_magnitudes_hst_hoix}. 
The image of the field around N5408X1 (Figure \ref{image}) highlights the presence of a small stellar association, close to the ULX at about 4\arcsec ($\sim 93\ \mathrm{pc}$) north-east with $\sim 20$ bright, resolved stars. In the field of view of our WFC3 observations, the other nearest association is at $\sim 8\arcsec$ north-west from the ULX, and is likely connected to the super star clusters that are located norther, as seen on previous broadband {\it HST} and SUBARU images \citep{Kaaret03,Soria06}.
This image also shows the nebula to be well resolved, with a diameter $\sim 2.5 \arcsec$ ($\sim 60\ \mathrm{pc}$), consistent with the size of the radio nebula \citep{Lang07,Cseh11b}. From optical long-slit spectroscopy, \citet{Kaaret09} derived a size of 30 pc (full width at half-maximum) for the \ion{He}{3} region along the slit. The size of the nebula in lower ionization emission lines, e.g., [\ion{O}{3}], is broader, about 40 pc (full width at half-maximum) with broad wings. This is quite consistent with the {\it HST} imaging.

\subsection{Stellar environment}

We used color-magnitude diagrams to investigate the stellar environment of the ULX. Data have been corrected by the reddening derived from the nebula surrounding the ULX, $E(\bv)=0.08$ \citep{Kaaret09} using the extinction law of \citet{Cardelli89} with $R_{\mathrm{V}} = 3.1$. The metallicity of NGC~5408 has been estimated to be about a tenth of the solar metallicity, with $12 + log(O/H) = 7.99$ \citep{Oliveira06}, which translates into $Z=0.002$ (based on a standard solar composition, \citealt{Grevesse98}).
We note that the age of the young and blue stars that are most common in the color-magnitude diagrams are not really sensitive to metallicity and therefore our age estimate does not depend on this parameter. However, the stars on the red supergiant branch are more sensitive to this effect; the red supergiants apparent in the (F547M,F845M) diagram indicate a subsolar metallicity, but with a best fit consistent with $Z=0.008$.

\begin{figure*}[tb]
\centerline{\resizebox{15cm}{!}{\includegraphics{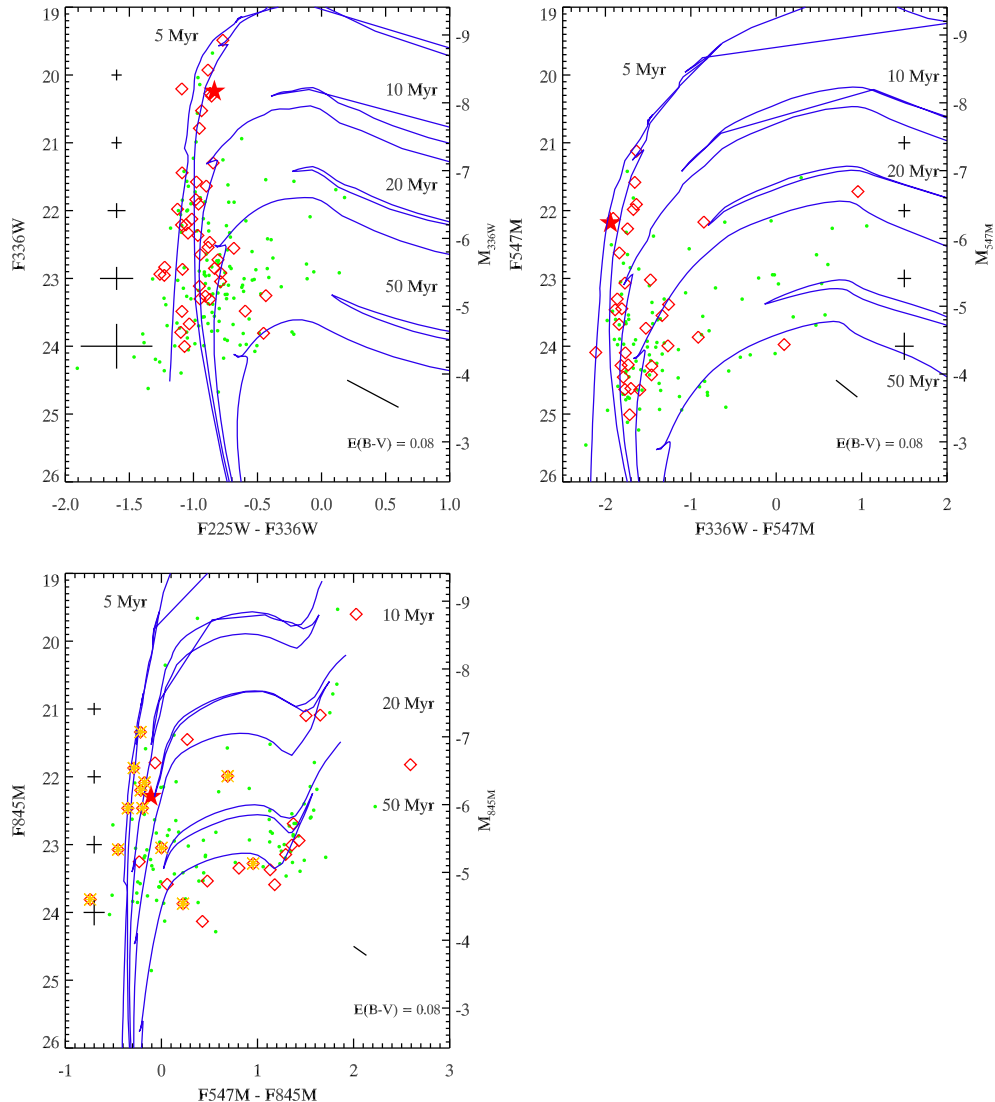}}}
\caption[]{{\it HST}/WFC3 color-magnitude diagrams for the stellar field ($512 \times 512\ \mathrm{pixels}$, i.e. $20\arcsec \times 20 \arcsec$)
   around the ULX. Padova isochrones for stars of different ages are 
   overplotted. Typical photometric errors are also plotted. Data 
   have been corrected for an extinction of E$(B-V)
   = 0.08$ mag, the bar at the bottom right corner illustrating this
   effect. 
   Upper left panel: Color-magnitude diagram in the (F225W,F336W) system.
   Upper right panel: Color-magnitude diagram in the (F336W,F547M) system.
   Lower left panel: Color-magnitude diagram in the (F547M,F845M) system. The same
   isochrones are plotted in the three panels, i.e for 5, 10, 20, and
   50 Myr at Z = 0.008. 
Filled green circles are stars in the field, red diamonds are stars located in a rectangle of $5.5 \times 4.0 \arcsec$ coincident with a nearby OB association. The red star is the position of the ULX counterpart. In the bottom panel, we show as orange X's stars from the OB association that are in common with the upper panels. \red{The smaller number of stars in common with the upper panels is simply due to the fact that we mainly see redder stars in the F845M filter} that are not strong UV emitters (and vice-versa in the top panels where the strong UV emitters are not seen in the redder filters).
}
\label{cmd}
\end{figure*}

\begin{figure*}[tb]
\centerline{\resizebox{15cm}{!}{\includegraphics{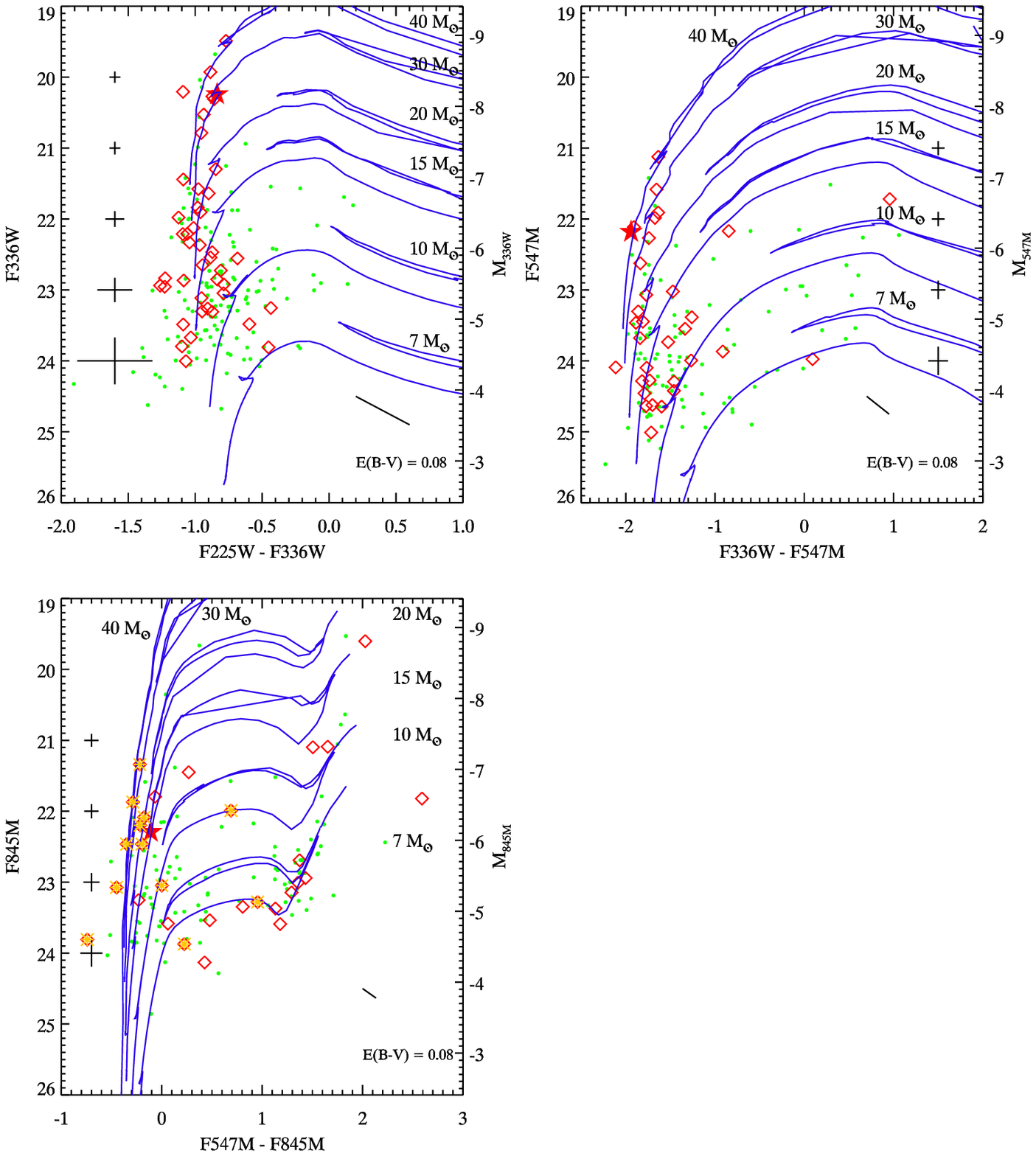}}}
   \caption[]{{\it HST}/WFC3 color-magnitude diagrams
   with Padova evolutionary tracks for stars of different initial masses
   ($7$, $10$, $15$, $20$, $30$, and $40\ \mathrm{M_{\sun}}$) with $Z=0.008$.\
   The legend is the same as Figure \ref{cmd}.
   Upper left panel: color-magnitude
   diagram in the $(F225W,F336W)$ system. Upper right panel: color-magnitude
   diagram in the $(F336W,F547M)$ system. Lower right panel: color-magnitude diagram in
   the $(F547M,F845M)$ system.
}
\label{et}
\end{figure*}

Thus, we overplotted isochrones from the Padova library \citep{Bertelli94,Marigo08,Girardi08,Girardi10} with a metallicity $Z=0.008$.
We present (Figure \ref{cmd}) three diagrams using the F225W, F336W, F547M and F845M filters. The diagrams with blue colors (upper panels of Figure \ref{cmd}) show that the stars associated with the relatively dense association nearby the ULX form a young upper main-sequence/supergiant branch that is well fitted by an isochrone of age $\sim 5\ \mathrm{Myr}$. A young age for the association is supported by the presence of six bright ($M_{F547M} \sim M_V = -6$ to $-7$) stars with blue colors that are most likely blue supergiants. Again, we emphasize that the age of these young stars does not depend much on the metallicity, using $Z=0.008$ instead of $Z=0.002$ introduces at most an uncertainty of 1 Myr which is smaller than the uncertainties due to photometric errors and isochrones models. We also overplotted Padova evolutionary tracks with various masses on the color-magnitude diagrams (Figure \ref{et}). From those, it is apparent that numerous stars from the association (and from the field) are massive objects with masses up to $30$--$40\ \mathrm{M_\odot}$.

\subsection{ULX}

We present (Figure \ref{opt_sed}) a detailed spectral energy distribution (SED) of this ULX, from 2400 to 15400~$\mathrm{\AA}$. The three SEDs taken at different epochs are well consistent with each other. In each case, we fitted the SED by using a simple power-law with $F_{\nu} \propto \nu^{\alpha}$, $F_\nu$ being the flux in units of $\mathrm{erg \; s^{-1} \; cm^{-2} \; Hz^{-1}}$. The indices of the power-law fits are respectively $1.37 \pm 0.03$, $1.33 \pm 0.03$, and $1.31 \pm 0.03$ for SED 1, 2 and 3. However, these fits are unacceptable with $\chi^2_{\nu}$ of 7.7, 7.7 and 5.0 for 4 degrees of freedom. Using only the F336W, F547M and F845M values to constrain the fit, we obtain a good fit with $\chi^2_{\nu} < 1$. The indices are not too different, with $1.38 \pm 0.08$, $1.40 \pm 0.07$, and $1.28 \pm 0.08$ for SED 1, 2 and 3 and the normalizations at $\lambda = 5500 \mathrm{\AA}$ are respectively $4.87 \pm 0.16$, $5.30 \pm 0.15$ and $5.54 \pm 0.17$ in units of $10^{-18}\ \mathrm{erg \; s^{-1} \; cm^{-2}}$ . This result is consistent with the index value found by \citet{Tao11} ($1.41 \pm 0.14$) from 4 simultaneous observations obtained by {\it HST}/WFPC2 in April 2009 and covering the same wavelength range ($3300$-$8000\ \mathrm{\AA}$).
As can be seen on Figure \ref{opt_sed}, those fits show a clear deviation from a straight power law in the NIR wavelengths, and also show marginal evidence for a turn-off at the shortest wavelengths. Interestingly, the deviations at these wavelengths seem to correspond to what is expected from a B0I star (Figure \ref{opt_sed}, right panel). Only the flux at 15400 \AA\ does not follow the decreasing trend that should continue in the infrared.
We note that the power-law index of the continuum in the 4000-6000~\AA\ interval found by \citet{Kaaret09} ($2.0^{+0.1}_{-0.2}$), is likely contaminated by the nebular emission. The spatial resolution of {\it HST}/WFC3 allows us to measure the flux of the continuum of the counterpart, with a very low contamination of the nebula as can be seen in the composite image (Figure \ref{image}) that contains the nebular [\ion{O}{3}] emission.

The three X-ray spectra observed simultaneously with the {\it HST} data have count rates of 0.30, 0.32 and 0.42 counts/s. Their moderate exposure times imply that the number of counts is limited ($\sim 4000$) and thus complex models such as used recently in the literature \citep{Gladstone09,Caballero10,Walton11} cannot be constrained accurately. Specifically, the lack of counts at high energy does not allow one to see the high-energy curvature as shown by \citet{Stobbart06}.
We modeled the X-ray spectra using basic models such as an absorbed power-law and a power-law plus disk blackbody ({\small DISKBB}, \citealt{Mitsuda84}) combination to compare with previous results in the literature. Extinction was modeled using the Tuebingen-Boulder ISM absorption model ({\small TBABS}, \citealt{Wilms00}). We used a fixed component related to the Galactic extinction with $n_\mathrm{H} = 0.057 \times 10^{22}\ \mathrm{cm^{-2}}$ \citep{Dickey90}, and let a second extinction component varying freely. N5408X1 does not appear to show any strange behavior. The fits parameters for these simple models (Table \ref{tab_fp}) are in agreement with previous observations \citep{Kajava09} where the authors argue that N5408X1 displays a $L_X$--$\Gamma$ correlation, which holds here.

To further understand the whole SED and the effect of X-ray emission at lower energies, we extrapolated the emission of the disk component. As shown by \citet{Kaaret04} in the case of Holmberg~II~X-1, some precautions need to be taken since the X-ray emission of most ULXs is largely dominated by the power-law component and not the accretion disk, and extrapolating the power law to lower energies would be non-physical. In the case of N5408X1, the ratio between the two components is about 50\% (Table \ref{tab_fp}), requiring care. Following \citet{Berghea10}, we modeled separately the powerlaw and the accretion disk component (Table \ref{tab_fp}). We fitted a powerlaw model to the spectra between 1 and 7 keV, and a disk blackbody model between 0.3 and 1 keV. As can be seen in Figure \ref{fit_xopt}, the extrapolated disk model underestimates the {\it HST} fluxes by about an order of magnitude, by a factor $\sim 38/36/24$ at 2400 \AA\ and by a factor $\sim 5/5/3$ at 15400 \AA\ .
We also used a model that consists of a disk plus a thermal Comptonization component ({\small DISKPN + COMPTT}, Table \ref{tab_fp}), as used in a recent study of N5408X1 \citep{Middleton11}. The comptonization component is a better choice since it is negligible in the optical wavelengths, so that there is no flux contamination in the disk component, as it is the case for the model based on a powerlaw. Using that model, we found that the extrapolated disk spectrum is 11/16/6 times lower at 2400 \AA\ and 5/8/3 times lower at 5400 \AA\ compared to {\it HST} fluxes. At 15400 \AA\, the extrapolated disk is 1.5/2.3 lower in the two first observations and slightly above for the third observation compared to {\it HST} fluxes (Figure \ref{fit_xopt}). We note that these extrapolations assume a very large outer disk radius \red{($\ga 1\times 10^{13}\ \mathrm{cm}$)}. Assuming narrower disks (or truncated disks) would reduce their flux in the IR, optical, and UV.

Reprocessed X-ray irradiation is likely to be significant in ULXs, the optical light curves of, e.g., NGC~1313~X-2 \citep{Grise08, Impiombato10} show short-term, stochastic variability that does not seem compatible with just the ellipsoidal variations due to a companion star.
Thus, following \citet{Kaaret09}, we fitted the X-ray and the {\it HST} data with the irradiated disk model of \citet{Gierlinski09}, {\small DISKIR}. It is based on the standard {\small DISKBB} model but includes effects due to disk irradiation and Comptonization. The irradiated inner disk and the comptonized tail illuminate the outer disk, which implies a higher luminosity for this part of the disk. This model has been applied successfully on a galactic black hole binary \citep{Gierlinski09} where the authors conclude that the optical/UV emission in the soft state is consistent with reprocessing of a constant fraction of the bolometric X-ray luminosity. It has also been applied to the microquasar GRS~1915+105 where the authors \citep{Rahoui10} show that an excess of mid-IR emission ($\sim 4-8 \mu m$) is probably related to some reprocessed soft X-ray emission in the outer part of the disk.

The {\small DISKIR} model contains nine parameters with two of them being constrained by the UV/optical/NIR measurements. Three parameters were frozen: {\it f$_{\mathrm{in}}$}, the fraction of luminosity in the Compton tail which is thermalized in the inner disk, was set to 0.1. The electron temperature and the irradiated radius were found to be poorly constrained so we fixed them respectively to 50 keV and $1.1 \times R_{\mathrm{in}}$. The electron temperature does not seem to have a strong impact on the other parameters. We tested with two other values (1.5 and 500 keV) but the effect on all parameters is within 20\%. Changing the irradiated radius from $1.1$ to $2.0 \times R_{\mathrm{in}}$ increases the inner temperature of the disk by 10\%, the ratio L$_{\mathrm{C}}$/L$_{\mathrm{D}}$ by 80\% and f$_{\mathrm{out}}$ by 20\% but this does not affect our conclusions. The results of the fits are shown in Table \ref{tab_diskir} and Figure \ref{fit_xopt} and are consistent for the three spectra. \red{We note that spectrum 1 has worse $\chi^2$ than the other spectra.  The residuals have positive peaks between 0.5 and 1~keV  and could be due to low energy emission lines. \citet{Caballero10} has suggested that ULXs may produce  highly ionized oxygen and iron emission lines via reflection off the disk. However, it is not clear why such lines would be present in the first spectrum only (that has the lowest number of counts). This interesting issue should be looked into with deeper spectra.  In this paper we limit ourselves to the continuum modeling.}

\section{Discussion}

\subsection {Environment of the ULX}
ULXs are usually located inside or near stellar-forming regions that look like OB associations \citep{Soria05,Grise08,Swartz09}. This has strenghtened their identification with X-ray binaries containing a massive companion star that is needed to explain their high, persistent X-ray luminosity.
The fact that N5408X1 is located nearby (but not inside) to what appears to be an OB association (Figure \ref{image}) is thus not unusual. For example, NGC~1313~X-2 \citep{Grise08} is displaced from its apparent parent cluster by $\sim 100\ \mathrm{pc}$. This is a good argument that such X-ray binaries may have been ejected from their host cluster \citep{Kaaret04b}.

\citet{Kaaret03} has shown that the ULX is located at about $12\arcsec$ from the star formation regions of NGC~5408 that contain super star clusters. The runaway-coalescence scenario studied by \citet{Vanbeveren09} (and references therein) is able to explain the formation of massive black holes up to several $100\ \mathrm{M_{\odot}}$ in subsolar metallicity environments. It is not clear if IMBHs could be produced by this way, but in any case the formation of such massive remnants would need massive and dense clusters. As noted by \citet{Kaaret03}, the projected displacement of $\sim 280$ pc from the super star clusters could be consistent with an ejected X-ray binary. Using reasonable ejection speed (10 km/s), \citet{Kaaret03} concluded that the binary could traverse this distance in 30 million years and thus have moved to the present location within the lifetime of a massive companion star. We remark here that 30 Myr would in fact exclude most early types of stars, since 30 Myr is the lifetime of a $\sim 10\ \mathrm{M_{\odot}}$ star.

Another possibility, as noted above, is that the binary comes from the closer OB association located at 100 pc from its position. In that case, it is very unlikely that an IMBH has been formed in such a stellar environment. We also note that the age of the association, $\sim 5\ \mathrm{Myr}$ would imply an escape velocity of $\sim 25-50\ \mathrm{km/s}$ with the assumption that the binary was ejected between 1 and 3 Myr after the stellar association was formed. \red{These velocities, albeit large, cannot be ruled out since some OB-supergiant binaries have been seen with runaway velocities of that order (e.g., \citealt{VandenHeuvel00}).} In that case, a more massive donor star would be allowed in the system, in comparison with the super star cluster hypothesis.

\subsection{UV/optical/NIR SED}

\begin{figure*}[tb]
\centerline{\includegraphics{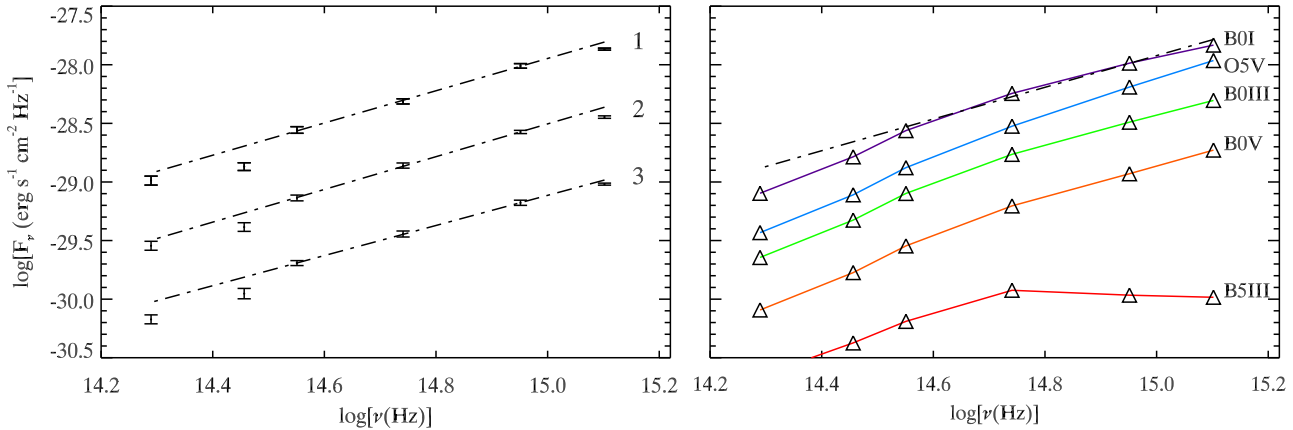}}
\caption[]{Spectral energy distribution of NGC~5408~X-1 (corrected for extinction) for each observation ({\it left}), compared to stellar templates ({\it right}). SEDs 2 and 3 of NGC~5408~X-1 have been shifted down by 0.6 in log($F_\nu$) for clarity purposes. Fluxes of stellar templates in the {\it HST}/WFC3 filters have been calculated using synphot and the Castelli and Kurucz stellar atmosphere models \citep{Castelli04}. In the right panel is overplotted (black dot dashed line) an average power law fit from the three HST observations.
}
\label{opt_sed}
\end{figure*}

The fact that the SED of N5408~X-1 may be described approximately by a power-law, and does not show any clear clue of a peak of a blackbody leads to several conclusions. As can be seen in Figure \ref{opt_sed}, only early type stars display a power-law extending far in the UV because the peak of their spectrum is located at shorter wavelengths than our observing limit ($\sim 2250\ \mathrm{\AA}$ equivalent to a peak temperature of $\sim 13000\ \mathrm{K}$). Basically, this means that if the light of N5408~X-1 was dominated by a star, we could rule out any star of spectral type later than B5 (this is quite independent of the luminosity class). But, the fact that the counterpart is intrinsically bright, $M_V \sim -6.2$ would exclude the smallest stars. Looking again at Figure \ref{opt_sed} (right panel) we see that the SED of N5408~X-1 is well consistent with that of a B0I supergiant star (scaled at the magnitude of a Iab class) and rules out all main-sequence stars. The turnoff in the F225W filter seems to match the spectral template of the B0I star. Also, a drop compared to a power law is expected in the NIR (Figure \ref{opt_sed}, right panel) and is clearly observed at $1 \mu m$ (Filter F105W). However, this drop should be even more prominent at longer wavelengths but is not really seen at $1.5 \mu m$ (Filter F160W).
Proving that the companion star dominates the optical emission would be possible if optical spectroscopy of the optical counterpart would reveal absorption lines, such as in the ULX P13 \citep{Motch10}. But the signal to noise of the optical spectra studied by \citet{Kaaret09} and \citet{Cseh11} are not high enough to rule out the presence of an early B supergiant star because absorption lines in the wavelength range observed are not very deep.

\subsection{X-ray to NIR SED}

\begin{figure*}
\centering
   \begin{tabular}{cc}
    \includegraphics[trim=6mm 2mm 0mm 0mm, clip, width=8.cm]{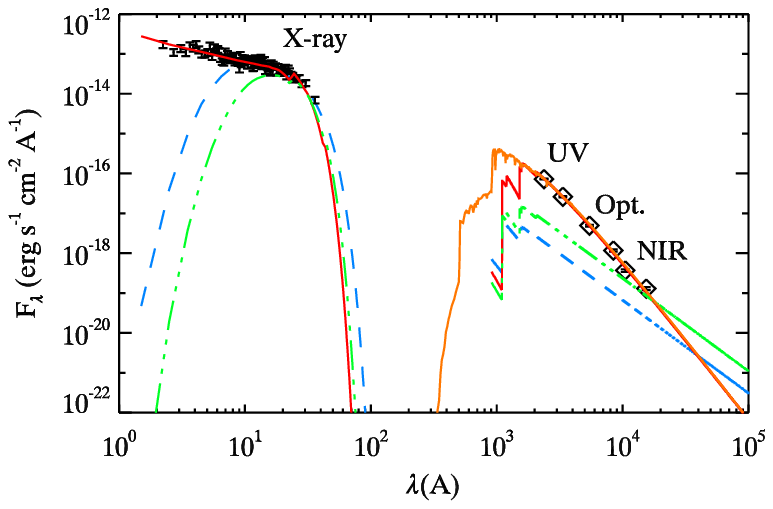}
   &\includegraphics[trim=6mm 2mm 0mm 0mm, clip, width=8.cm]{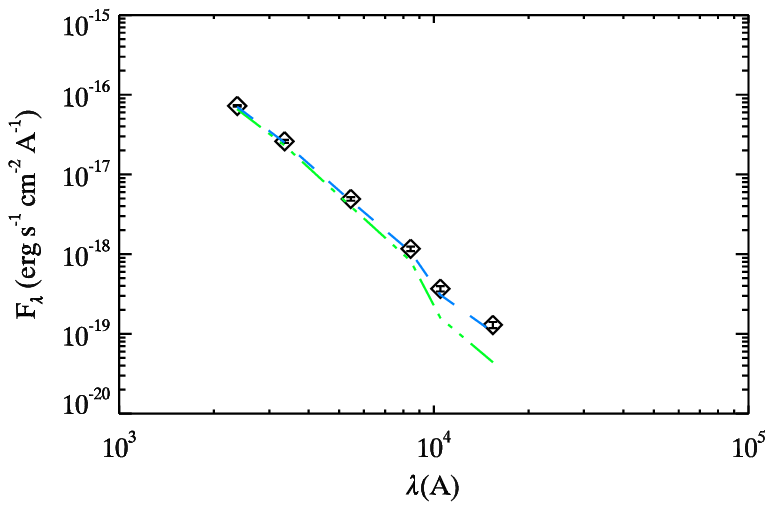}\\
    \includegraphics[trim=6mm 2mm 0mm 0mm, clip, width=8.cm]{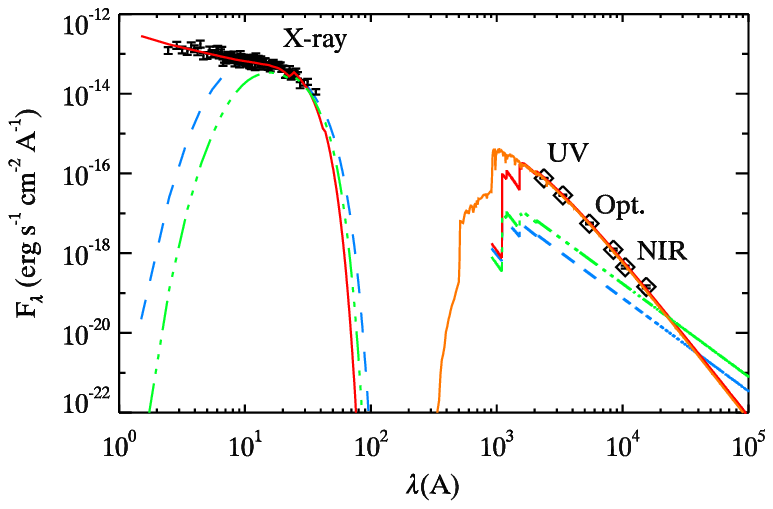}
   &\includegraphics[trim=6mm 2mm 0mm 0mm, clip, width=8.cm]{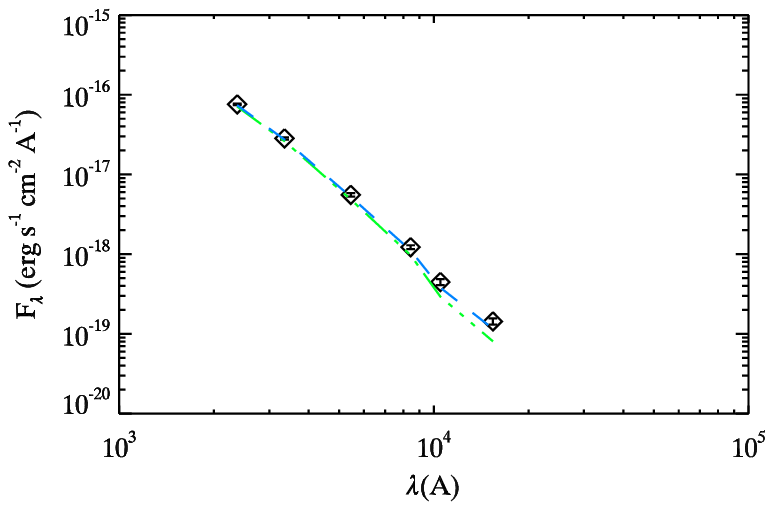}\\
    \includegraphics[trim=6mm 2mm 0mm 0mm, clip, width=8.cm]{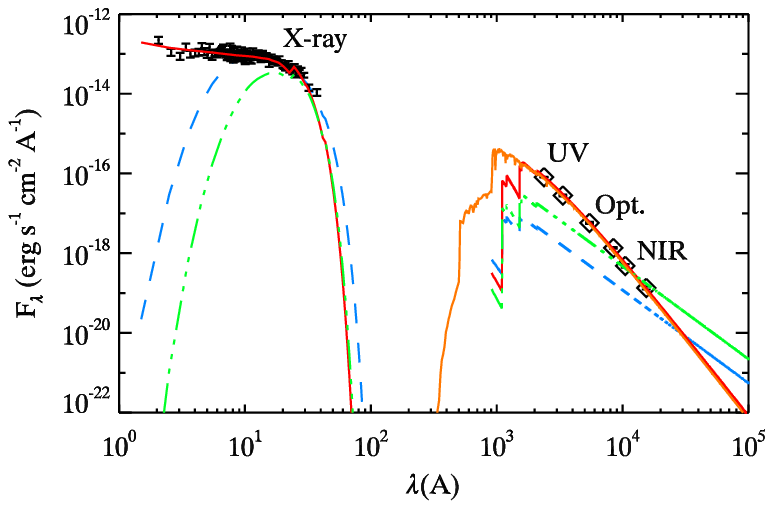}
   &\includegraphics[trim=6mm 2mm 0mm 0mm, clip, width=8.cm]{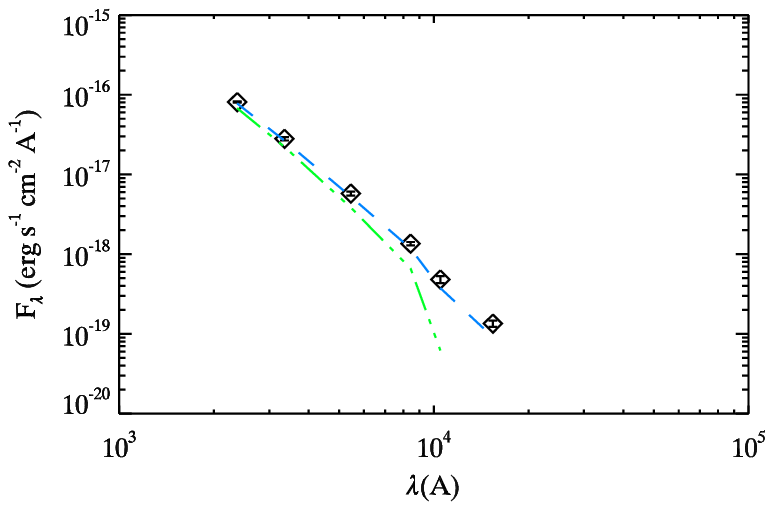}\\
    \end{tabular}
   \caption[]{Left: best fits for the simultaneous X-ray - UV/optical/NIR data of NGC~5408~X-1. The three plots correspond to the three independent observations. In each plot, the two accretion disk models (see text) extrapolated from the X-ray data fit to lower energies are shown in green (dot-dashed line) and in blue (dashed line)\red{, respectively corresponding to the disk model associated with a Comptonized component and the disk model associated with a power law component}. They underestimate the UV/optical/NIR fluxes.
The best fitting model is an irradiated disk (red line) that explains both the high energy and low energy data. This supports the interpretation that a standard, thin disk is present in NGC~5408~X-1. The spectrum of a B0I supergiant star (from the Castelli and Kurucz stellar atmosphere models, Castelli et al. 2004) is overplotted in orange.
It fits the {\it HST} data as well as the irradiated disk model. Note that the apparent gap between $\sim 100$ and $\sim 1000\ \mathrm{\AA}$ is due to the extinction. Right: required flux needed to explain the {\it HST} measurements, after subtraction of \red{the two extrapolated disk components seen in the left panels (the same color/linestyle is used to differentiate the two disk models)}.
} 
   \label{fit_xopt}
\end{figure*}

The disk emission, extrapolated from two different models that we used to fit the X-ray data (a disk blackbody with the addition of either a power-law or a thermal Comptonization model) does not fit the UV/optical/IR at all (Figure \ref{fit_xopt}). For the power-law plus disk blackbody model, the extrapolated disk model underestimates the {\it HST} fluxes by about an order of magnitude (Figure \ref{fit_xopt}) which basically means that the accretion disk emission would be negligible at optical wavelengths and the companion star would dominate.

For the disk plus Comptonization model, the extrapolated disk spectrum has a higher flux at lower energies but is still unable to explain the UV/optical/IR measurements (Figure \ref{fit_xopt}). The excess emission could come from the donor star. However, the shape of the excess emission drops at the longest wavelengths, especially in the third observation where the NIR flux predicted by the accretion disk model is above the flux measured with {\it HST} (Figure \ref{fit_xopt}). This is not consistent with the spectrum of any known star. 

Another possibility is that the emission in excess from the extrapolation of the disk may not come primarily from a stellar component. The irradiated disk model that we used is able to fit all the data quite well (see Table \ref{tab_diskir}). All three datasets give similar results within the errors, which is expected given the lack of significant variability both at X-rays and at lower energies. In this model, the UV/optical/IR emission is from the irradiated outer disk that thermalizes a fraction of the bolometric luminosity $f_{out} \sim 0.03$. This fraction is higher than seen from Galactic X-ray binaries in the thermal dominant state \citep{Gierlinski09, Zurita11} but similar fractions have been obtained from observations of the microquasar GRS1915+105 in its most X-ray active phase \citep{Rahoui10}.

\begin{table*}
\centering
\begin{threeparttable}
\caption[]{Spectral Fit Parameters of the Irradiated Disk Model}
\centering
\tiny
\begin{tabular}{l*{10}{c}r}
\hline
No.\tnote{a} & $n_\mathrm{H}$\tnote{b}  & diskir$_{norm}$\tnote{c} & $\Gamma$\tnote{d} & $kT_{\mathrm{in}}$\tnote{e} & kT$_e$\tnote{f} & L$_C$/L$_D$\tnote{g} & r$_{irr}$\tnote{h} & f$_{out}$\tnote{i} & log (r$_{out}$)\tnote{j} & $\chi^2$/DoF\tnote{k}\\
     		     & ($10^{22}\ \mathrm{cm^{-2}}$) &   &  & (keV)  & (keV) &&&&&\\
\hline
\multicolumn{11}{c}{DISKIR} \\
\hline
1	& 0.16$^{+0.06}_{-0.05}$ & 666$^{+21}_{-470}$	& $2.42^{+0.16}_{-0.17}$   & 0.13$^{+0.02}_{-0.02}$ & 50 & 0.57$^{+0.56}_{-0.28}$ & 1.1  & 0.043$^{+0.030}_{-0.022}$  & 3.04$^{+0.27}_{-0.25}$ &   144.6 (93)  \\
2	& 0.12$^{+0.08}_{-0.05}$ & 401$^{+1}_{-285}$	& $2.40^{+0.17}_{-0.19}$   & 0.14$^{+0.02}_{-0.02}$ & 50 & 0.71$^{+0.66}_{-0.37}$ & 1.1  & 0.046$^{+0.027}_{-0.024}$  & 3.21$^{+0.28}_{-0.28}$ &   101.7 (96)  \\
3	& 0.16$^{+0.07}_{-0.05}$ & 794$^{+34}_{-557}$	& $2.82^{+0.18}_{-0.19}$   & 0.14$^{+0.02}_{-0.02}$ & 50 & 0.43$^{+0.25}_{-0.20}$ & 1.1  & 0.030$^{+0.019}_{-0.018}$  & 3.07$^{+0.27}_{-0.29}$ &   114.0 (103)  \\
\hline
\hline
\end{tabular}
\label{tab_diskir}
\begin{tablenotes}
\footnotesize
\item[a] Spectrum index used in the text
\item[b] Total absorption column, including the galactic extinction towards the source ($n_\mathrm{H} = 0.057 \times 10^{22}\ \mathrm{cm^{-2}}$)
\item[c] Model normalization
\item[d] Power-law photon index
\item[e] Innermost temperature of the unilluminated disk
\item[f] Electron temperature
\item[g] Ratio of luminosity in the Compton tail to that of the unilluminated disk
\item[h] Radius of the Compton illuminated disk in terms of the inner disk radius
\item[i] Fraction of bolometric flux which is thermalized in the outer disk
\item[j] log10 of the outer disk radius in terms of the inner disk radius
\item[k] $\chi^2$ and degrees of freedom
\item
\item All errors are at the 90\% confidence level.
\end{tablenotes}
\end{threeparttable}
\end{table*}

\citet{Cseh11} have discussed the size of the accretion disk in N5408X1, based on the broad emission lines seen in the optical spectrum of the counterpart. They concluded that the disk has a radius lower than $2.35 \times \mathrm{M_{BH}/1500\ M_{\odot}\ AU}$ which translates to $2.4 \times 10^{11}$--$10^{13}\ \mathrm{cm}$ for a $10$--$1000\ \mathrm{M_{\odot}}$ black hole. According to the results from the irradiated disk model, we found that the disk would have a size $\sim 1-5 \times 10^{12}\ \mathrm{cm}$ depending on the inclination of the disk. This is comparable to the size of the disk inferred in GRS1915+105 with $\sim 3\times 10^{12}\ \mathrm{cm}$, where an irradiated accretion disk may also be present \citep{Rahoui10}.
In N5408X1, the irradiated disk model gives $R_{\mathrm{out}}/R_{\mathrm{in}} \sim 1100$--$1600$ which is in the upper range of other ULXs, like in NGC~1313~X-1 \citep{Yang11} with $R_{\mathrm{out}}/R_{\mathrm{in}} \sim 100$--$2000$ and in NGC 6946 X-1 \citep{Kaaret10} with $R_{\mathrm{out}}/R_{\mathrm{in}} \sim 40$--$6000$.

While disk irradiation is able to explain the observational properties of N5408X1, we cannot make a definitive statement that irradiation is occurring in this system. First, no detection of significant, associated variability between the UV/Optical/NIR and X-ray has been made (see below). Second, it is possible that the X-ray soft component is not emitted by an accretion disk, but by the photosphere of an outflow/wind (e.g., \citealt{Poutanen07, Middleton11}) that would be due to supercritical accretion onto the accretion disk.

In that context, it may be interesting to draw a comparison with the X-ray transient V4641 Sgr. This is an interesting source because it may be a closer example to ULXs than low-mass X-ray binaries (LMXBs) thanks to its companion that is a quite massive star ($M \sim 6.5\ \mathrm{M_{\mathrm{\odot}}}$). V4641 Sgr went into outburst in 1999 reaching an X-ray luminosity near or above the Eddington limit of its $\sim 10\ \mathrm{M_{\mathrm{\odot}}}$ black hole \citep{Revnivtsev02a}. Along with the X-ray outburst, the optical emission increased by at least 4.7 mag. It has been shown that irradiation in the accretion disk cannot be reconciliated with the observed parameters. Instead, \citet{Revnivtsev02a,Revnivtsev02b} argue that an extended envelope surrounding the source absorbs the X-ray flux and reemits it in the optical and UV. This would be the signature of a massive outflow driven by the supercritical accretion. The optical nebulae present around some ULXs and around N5408X1 (Figure \ref{image}) argue for the presence of such outflows. As noted in \citet{Kaaret09} the difficulty with the supercritical model is that the soft X-rays expected from the photosphere of the wind would not thermalize in the disk and thus would not contribute to the reprocessed emission in the optical bands. This is a natural consequence of the supercritical accretion models because they predict geometric beaming \citep{Poutanen07, Ohsuga11}. If the irradiated disk model is valid, it means that such beaming can be ruled out in N5408X1, because of the high fraction of thermalized bolometric luminosity in the disk. This would challenge the supercritical model. On the contrary, if the optical spectrum is dominated by the donor emission, then mild geometric beaming consistent with the constraints from the surrounding \ion{He}{2}4686 nebula would be allowed. Thus confirming or not that the optical emission comes mainly from the donor star is important.

\red{We finally note that the reflection-based model used to model the X-ray spectrum of N5408X1 \citep{Caballero10} would probably imply a negligible contribution of the  accretion disk in the optical wavelengths. This is because there is apparently no need for thermal emission from the disk in that model. This may be explained by magnetic extraction of energy from the disk, that would suggest that the energy is extracted before it is able to thermalize in the disk \citep{Caballero10}). In that context, irradiation in the outer parts of the disk would be at best limited, and the observed optical emission would be mainly due to the donor star.}

\subsection{Variability}

Observing X-ray/optical correlated variability would be a direct test of the irradiated disk model. A lack of significant optical variability in response to a significant X-ray variation would rule out the model, and would probably argue for the dominance of the companion star at non-X-ray wavelengths.
The X-ray variability we measure is limited, with a 40\% change in the count rate between the first and last observation, and with the second observation displaying a flux consistent with that of the first one. At longer wavelengths, there is no clear, statistically significant variability in the {\it HST} filters. In practically all filters, the magnitudes in all three epochs are consistent within the errors (Table \ref{tab_magnitudes_hst_hoix}), with only marginal evidence of variability between the first and third epoch, at a level $\la 2 \sigma$.
Thus, we can only provide an upper limit on the variability in the {\it HST} filters, consistent with $\la 0.1\ \mathrm{mag}$ in most filters. If the situation is comparable as that of LMXBs, we would expect the amplitude of optical variability to be roughly the square root of the amplitude of the X-ray variability \red{\citep{VanParadijs94}}. Our results are not inconsistent with this because the X-ray variability would imply a change of the optical flux of only $\sim$18\%, which is close enough to 10\% and also because other processes can affect the optical luminosity of the system with approximately the same amplitude (i.e., ellipsoidal variations).

\begin{figure}[tb]
\centerline{\resizebox{8cm}{!}{\includegraphics{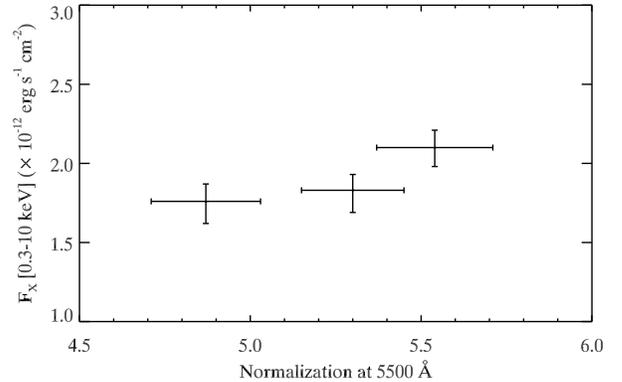}}}
   \caption[]{Absorbed X-ray flux (0.3-10 keV) versus normalization at 5500 \AA. X-ray fluxes are taken from the power law fits and normalizations come from the F336W/F547M/F845M fits.
}
\label{corr_fx_fopt}
\end{figure}

Perhaps the best evidence of variability may be seen in the normalization of the power-law fits. At 5500 \AA\ the difference in the normalization factors between the first and last observation is $0.67 \pm 0.23\ \times 10^{-18}\ \mathrm{erg\ s^{-1}\ cm^{-2}}$ which is almost significant at a $3 \sigma$ level. We tried to look for a correlation between the X-ray flux and the magnitude/normalization in all filters but the errors on both parameters are too large (see Figure \ref{corr_fx_fopt}).

The long term X-ray light curve of N5408~X-1 measured with {\it Swift} shows dips quite regularly \citep{Kong11}, with the X-ray luminosity decreasing by nearly an order of magnitude compared to the long-term average. In the context of an irradiated disk, this would imply a drop in the optical flux by a factor $\sim 3$, i.e. $\sim 1\ \mathrm{mag}$ and thus would be easily measurable. \citet{Gierlinski09} noted that this 'reprocessing signature' is approximate, depending on parameters such as the size of the disk, but given the order of magnitude of the variability that we expect, it should still be detectable, even if reduced, in the UV/Optical/NIR SED of the source.
Finally, another additional signature would be an associated decrease in the flux of the \ion{He}{2} emission line visible in the optical spectrum of the source \citep{Cseh11}, if the emission of \ion{He}{2} is really associated with reprocessing in the accretion disk. Thus, it would be of interest to perform optical observations during an X-ray dip.

\section{Conclusion}

We have studied NGC~5408~X-1 using 3 epochs of simultaneous {\it Chandra}/{\it HST} observations. The optical counterpart of the ULX is visible from the UV to the NIR. The {\it HST}/WFC3 observations reveal that the souce is located near a young, $\sim$ 5 Myr old, OB association containing massive stars up to $40\ \mathrm{M_{\odot}}$, with a few blue supergiants. This could be the birthplace of the ULX system, but it cannot be ruled out that it was formed instead in a superstar cluster in NGC~5408.

The UV to NIR SED is compatible with that of a B0I supergiant star. Using the simultaneous X-ray data, we show that the intrinsic emission of a standard accretion disk cannot explain the UV/Optical/NIR fluxes. Instead, a model that takes into account irradiation in the disk fits well all the data. Further test of this model requires observing the source with a lower or higher X-ray flux and studying the correlation with the optical flux. This would be a good discriminant toward determining the physical origin of the optical emission.

\section*{Acknowledgments}

We thank Leo Girardi for making available the Padova evolutionary tracks in the HST/WFC3 medium filters. FG and PK acknowledge support from Chandra grant GO0-11050 and STScI grant HST-12021.  SC and DC are grateful for support from the European Community's Seventh Framework Program (FP7/2007-2013) under grant agreement number ITN 215212 "Black Hole Universe". We thank the anonymous referee for their comments which improved this paper.



\begin{sidewaystable}[htp]
\begin{threeparttable}
\caption[]{Spectral Fit Parameters}
\centering
\tiny
\begin{tabular}{l*{10}{c}r}
\hline
No.\tnote{a}  & $n_\mathrm{H}$\tnote{b}  & $\Gamma$\tnote{c} (kT$_e$)\tnote{d} & $\Gamma_{norm}$\tnote{e} (Compt. norm)\tnote{f} & $kT_{\mathrm{in}}$\tnote{g} & Disk$_{norm}$\tnote{h} & $\tau$\tnote{i} & Flux\tnote{j} & $L_X$\tnote{k} & $f_{X_{\mathrm{MCD}}}$\tnote{l} & $\chi^2$/DoF\tnote{m}\\
     	& ($10^{22}\ \mathrm{cm^{-2}}$) & ( /keV)  & $10^{-4}$  &   &  &   & ($\times 10^{-12}\ \mathrm{erg\ s^{-1}\ cm^{-2}}$) & ($\times 10^{40}\ \mathrm{erg\ s^{-1}}$) & &\\
\hline
\multicolumn{10}{c}{POWERLAW} \\
\hline
1	&0.044$^{+0.014}_{-0.014}$	     & 2.66$^{+0.08}_{-0.08}$	& $5.8^{+0.3}_{-0.3}$		      & ...& ... & ... & 1.76$^{+0.11}_{-0.14}$ & 0.63 $^{+0.04}_{-0.04}$  & ...  & 154.36 (91) \\
2	&0.03$^{+0.02}_{-0.02}$	     & 2.67$^{+0.13}_{-0.12}$	& $5.9^{+0.5}_{-0.4}$		      & ...& ... & ... & 1.83$^{+0.10}_{-0.14}$ & 0.79 $^{+0.08}_{-0.05}$  & ...  & 121.0 (94) \\
3	&0.10$^{+0.02}_{-0.02}$	     & 3.17$^{+0.13}_{-0.12}$	& $9.9^{+0.2}_{-0.7}$		      & ...& ... & ... & 2.10$^{+0.11}_{-0.12}$ & 1.54 $^{+0.23}_{-0.17}$  & ...  & 136.5 (101) \\
\hline
\hline
\multicolumn{10}{c}{POWERLAW + DISKBB} \\
\hline
1	 &0.10$^{+0.05}_{-0.04}$	     & 2.43$^{+0.12}_{-0.12}$ 	& 4.9$^{+0.6}_{-0.6}$  		      & 0.15 $^{+0.02}_{-0.02}$ & 284$^{+577}_{-187}$ & ... & 1.76$^{+0.15}_{-0.76}$ & 0.85$^{+0.44}_{-0.19}$  & 0.28$^{+0.44}_{-0.18}$  & 132.03 (89) \\
2	 &0.07$^{+0.07}_{-0.05}$	     & 2.40$^{+0.18}_{-0.19}$ 	& 4.7$^{+0.9}_{-0.9}$ 		      & 0.16 $^{+0.04}_{-0.03}$ & 162$^{+809}_{-129}$ & ... & 1.86$^{+0.1}_{-0.76}$ & 1.02$^{+0.56}_{-0.25}$  & 0.36$^{+0.51}_{-0.20}$  & 100.6 (92) \\
3	 &0.13$^{+0.05}_{-0.05}$	     & 2.85$^{+0.18}_{-0.20}$ 	& 1.6$^{+1.5}_{-1.5}$ 		      & 0.15 $^{+0.04}_{-0.02}$ & 427$^{+1614}_{-327}$ & ... & 2.15$^{+0.31}_{-0.72}$ & 2.13$^{+1.25}_{-0.62}$  & 0.48$^{+0.52}_{-0.26}$  & 110.7 (99) \\
\hline
\hline
\multicolumn{10}{c}{Modified POWERLAW + DISKBB} \\
\hline
1	 &0.01$^{+0.04}_{-0.01}$	     & 2.51$^{+0.11}_{-0.10}$ 	& 5.0$^{+0.3}_{-0.3}$  		      & 0.27 $^{+0.04}_{-0.05}$ & 18$^{+48}_{-9}$ & ... & ... & ...  & ...  & 85.7 (54) / 59.8 (38) \\
2	 &0.0$^{+0.06}_{-0.0}$	     & 2.51$^{+0.11}_{-0.10}$ 	& 5.2$^{+0.6}_{-0.3}$ 		      & 0.26 $^{+0.02}_{-0.02}$ & 23$^{+34}_{-7}$ & ... & ... & ...  & ...  & 61.7 (55) / 37.3 (39) \\
3	 &0.02$^{+0.05}_{-0.02}$	     & 2.92$^{+0.10}_{-0.10}$ 	& 7.9$^{+0.4}_{-0.4}$ 		      & 0.26 $^{+0.04}_{-0.04}$ & 39$^{+80}_{-23}$ & ... & ... & ...  & ...  & 69.9 (62) / 44.6 (40) \\
\hline
\hline
\multicolumn{9}{c}{DISKPN + COMPTT} \\
\hline
1	& 0.06$^{+0.06}_{-0.05}$	     & 1.1$^{+37.2}_{-0.3}$ 	& 9.1$^{+3.8}_{-9.1}$ 		      & 0.15 $^{+0.04}_{-0.03}$ &  $3.5^{+12.2}_{-2.5}$  & 9.2$^{+2.8}_{-9.2}$ & 1.69$^{+0.11}_{-0.05}$ & 0.60$^{+0.33}_{-0.03}$ & 0.42$^{+0.37}_{-0.15}$  & 128.5 (88) \\
2	& 0.03$^{+0.06}_{-0.03}$	     & 1.0$^{+3.5}_{-0.3}$ 		& 8.3$^{+3.7}_{-2.8}$ 		      & 0.17 $^{+0.04}_{-0.03}$ & 2.0$^{+6.9}_{-1.2}$ & 10.2$^{+2.2}_{-7.3}$ & 1.80$^{+0.1}_{-1.4}$ &0.59$^{+0.28}_{-0.03}$ & 0.43$^{+0.57}_{-0.15}$  & 94.8 (91) \\
3	& 0.09$^{+0.07}_{-0.06}$	     & 9.8$^{+17.2}_{-8.6}$ 	& 2.4$^{+0.3}_{-2.4}$ 		      & 0.14 $^{+0.03}_{-0.02}$ & 10.1$^{+18.0}_{-7.0}$ & 1.4$^{+5.8}_{-1.4}$& 2.14$^{+0.35}_{-0.71}$ & 1.73$^{+1.08}_{-0.56}$ & 0.56$^{+0.44}_{-0.32}$  & 109.2 (98) \\
\end{tabular}
\label{tab_fp}
\begin{tablenotes}
\footnotesize
\item[a] Spectrum index used in the text
\item[b] External absorption column (in addition of the galactic extinction towards the source, $n_\mathrm{H} = 0.057 \times 10^{22}\ \mathrm{cm^{-2}}$)
\item[c] Power-law photon index
\item[d] Electron temperature in keV, for the Comptonization models
\item[e] Power-law normalization
\item[f] Comptonization normalization
\item[g] Inner-disk temperature
\item[h] Disk normalization (for the diskpn model, units are in terms of $10^{-3}$)
\item[i] Plasma optical depth
\item[j] Absorbed flux (0.3--10 keV) 
\item[k] Unabsorbed luminosity (0.3--10 keV) for $D=4.8\ \mathrm{Mpc}$
\item[l] Fraction of the total unabsorbed flux (0.3--10 keV) in the disk component
\item[m] $\chi^2$ and degrees of freedom
\item
\item All errors are at the 90\% confidence level.
\end{tablenotes}
\end{threeparttable}
\end{sidewaystable}

\end{document}